\documentclass[english]{article}
\usepackage[T1]{fontenc}
\usepackage[latin9]{inputenc}
\usepackage{amsmath}

\makeatletter

\providecommand{\tabularnewline}{\\}

\makeatother

\usepackage{babel}
\begin{document}

\title{Gamow-Teller transitions and magnetic moments using various interactions}

\author{Ricardo Garcia and Larry Zamick\\
Department of Physics and Astronomy, Rutgers University, Piscataway,
New Jersey 08854 \\
}
\maketitle
\begin{abstract}
In a single j-shell calculation we consider the effects of several
different interactions on the values of Gamow-Teller (B(GT)'s) and
magnetic moments. The interactions used are MBZE, J=0 pairing, J$_{max}$
pairing and half and half. Care is taken when there are isospin crossings
and/or degeneracies.
\end{abstract}

\section{Introduction}

In examining the spectrum of a system of a neutron and a proton beyond
a closed shell one sees that not only the J=0 T=1 but also J=1 T=0
and J=J$_{max}$=2j lie low. For example in $^{42}$Sc the matrix
elements taken from experiment by Escuderos et al. {[}1{]} are shown
in Table I:

\begin{center}
\textbf{Table I: Experimental two-body matrix elements}
\par\end{center}

\begin{center}
\begin{tabular}{|c|c|c|c|}
\hline 
T=1 & E & T=0 & E\tabularnewline
\hline 
\hline 
J &  & J & \tabularnewline
\hline 
0 & 0.0000 & 1 & 0.6111\tabularnewline
\hline 
2 & 1.5865 & 3 & 1.4904\tabularnewline
\hline 
4 & 2.8135 & 5 & 1.5101\tabularnewline
\hline 
6 & 3.2420 & 7 & 0.6163\tabularnewline
\hline 
\end{tabular}
\par\end{center}

In this work we will consider the above interaction which we call
MBZE, as well as some exteme interactions.

a. J=0 pairing: the 8 matrix elements are respectively -1,0,0,0,0,0,0,0

b. J$_{max}$ pairing: 0,0,0,0,0,0,0,-1

c. Half and half: -1,0,0,0,0,0,0,-1.

We will study how Gamow-Teller B(GT) values and magnetic moments in
the f$_{7/2}$ shell respond to these different interactions.

\section{Gamow-Teller B(GT) values}

We start with the well known formula for the case where the Fermi
matrix element vanishes.

\begin{center}
$ft=6177/[B(F)+1.583/B(GT)]$
\par\end{center}

In an allowed Fermi transition neither the total anguar momentum nor
the isospin can change. We will only consider cases where one or both
change so that B(F)=0.

We then obtain 

\begin{center}
$ft=3902.0846497/B(GT)$
\par\end{center}

\begin{center}
$log(ft)=3.591266854-log(B(GT))$
\par\end{center}

We will be using bare operators throughout.

As an orientation we note that for a free neutron B(GT)=3.

With the interactions mentioned in the introduction we can go to more
complex systems and obtain wave functions that are represented by
amplitudes D$^{I}$(J$_{p}$, J$_{n}$). The square of this amplitude
is the probability that in a state I the protons couple to J$_{p}$
and the neutrons to J$_{n}$.

We first consider a simple case where we do not require the amplitude
of the transition $^{42}$Sc (I=7$^{+}$) $\rightarrow$ $^{42}$Ca
(I=6$^{+}$). The initial state has isospin T=0 and the final T=1.

The experimental value is B(GT)=0.2699, while the theoretical value,
assuming a configuration (f$_{7/2}$)$^{2}$ for both the initial
and final states, is 0.2743. Thus, to agree with experiment, one needs
a quenching factor of 0.992 for the GT operator. In ref {[}2{]} this
quenching factor was used. However, in this third work we will stick
with the bare operator. It is worth mentioning that in this case we
have a proton changing into a neutron inside the nucleus and a positron
and neutrino escaping.

We now show results in Table II which do depend on the amplitudes.
The expression for B(GT) is given in 2 previous publications and is
here repeated.

\begin{center}
$X_{1}=\sum_{J_{p},J_{n}}D^{f}(J_{p},J_{n})D^{^{i}}(J_{p},J_{n})U(1J_{p}I_{f}J_{n};J_{p}I_{i})\sqrt{J_{p}(J_{p}+1)}$
\par\end{center}

\begin{center}
$X_{2}=\sum_{J_{p},J_{n}}D^{f}(J_{p},J_{n})D^{^{i}}(J_{p},J_{n})U(1J_{n}I_{f}J_{p};J_{n}I_{i})\sqrt{J_{n}(J_{n}+1)}$
\par\end{center}

\begin{center}
$B(GT)=\frac{1}{2}\frac{2I_{f}+1}{2I_{i}+1}f(j)^{2}\left[\frac{\left\langle 1T_{i}1M_{T_{i}}|T_{f}M_{T_{f}}\right\rangle }{\left\langle 1T_{i}0M_{T_{i}}|T_{f}M_{T_{i}}\right\rangle }\right]^{2}(X_{1}-(-1)^{I_{f}-I_{i}}X_{2})^{2}$
\par\end{center}

where 

\begin{center}
$f(j)=\begin{cases}
\frac{1}{j} & j=l+\frac{1}{2}\\
\frac{-1}{j+1} & j=l-\frac{1}{2}
\end{cases}$
\par\end{center}

If $T_{f}\neq T_{i}$ or $I_{f}\neq I_{i}$, we find that $X_{1}=-(-1)^{I_{f}-I_{i}}X_{2}$.
We then get a simplified formula for B(GT):

\begin{center}
$B(GT)=2\frac{2I_{f}+1}{2I_{i}+1}f(j)^{2}\left[\frac{\left\langle 1T_{i}1M_{T_{i}}|T_{f}M_{T_{f}}\right\rangle }{\left\langle 1T_{i}0M_{T_{i}}|T_{f}M_{T_{i}}\right\rangle }\right]^{2}(X_{1})^{2}$
\par\end{center}

This formula does not apply to the case of neutron decay because in
that case, $I_{f}=I_{i}$ and $T_{f}=T_{i}$. 

\begin{center}
\textbf{Table II: B(GT) values}
\par\end{center}

\begin{center}
\begin{tabular}{|c|c|c|c|c|c|c|c|}
\hline 
Transition & $I_{i}$ & $I_{f}$ & MBZE & J=0 & J=Half & J=7 & Experiment\tabularnewline
\hline 
\hline 
$^{43}$Sc $\rightarrow^{43}$Ca & 3.5 & 2.5 & 0.1181 & 0 & 0.0592 & 0.2434 & 0.0326\tabularnewline
\hline 
$^{43}$Sc $\rightarrow^{43}$Ca & 3.5 & 3.5 & 0.1682 & 0.5713 & 0.2747 & 0.0397 & \tabularnewline
\hline 
$^{43}$Sc $\rightarrow^{43}$Ca & 3.5 & 4.5 & 8.31$\times$10$^{-6}$ & 0 & 3.29$\times$10$^{-4}$ & 0.00136 & \tabularnewline
\hline 
$^{44}$Sc $\rightarrow^{44}$Ca & 2 & 2 & 0.0505 & 0.0613 & 0.0142 & 0.0259 & 0.01962\tabularnewline
\hline 
$^{45}$Sc $\rightarrow^{45}$Ca & 3.5 & 2.5 & 0.0094 & 0 & 0.0094 & 2.32$\times$10$^{-5}$ & \tabularnewline
\hline 
$^{45}$Ca $\rightarrow^{45}$Sc & 3.5 & 3.5 & 0.0552 & 0.4571 & 0.1423 & 4.49$\times$10$^{-4}$ & \tabularnewline
\hline 
$^{45}$Sc $\rightarrow^{45}$Ca & 3.5 & 4.5 & 1.64$\times$10$^{-4}$ & 0 & 3.16$\times$10$^{-4}$ & 1.03$\times$10$^{-5}$ & \tabularnewline
\hline 
$^{45}$Ti $\rightarrow^{45}$Sc & 3.5 & 3.5 & 0.1466 & 0.1499 & 0.1732 & 5.89$\times$10$^{-4}$ & 0.0980\tabularnewline
\hline 
$^{46}$Ti $\rightarrow^{46}$V & 4 & 4 & 0.0065 & 0.0166 & 0.2898 & 2.03$\times$10$^{-4}$ & 0.0025\tabularnewline
\hline 
$^{46}$Ti $\rightarrow^{46}$V & 4{*} & 4 & 0.0058 & 0.5458 & 0.0018 & 6.36$\times$10$^{-4}$ & 0.0025\tabularnewline
\hline 
$^{46}$Ti $\rightarrow^{46}$V & 1 & 0 & 0.0789 & 0 & 0.0367 & 0.2332 & 0.0196\tabularnewline
\hline 
$^{46}$Ti $\rightarrow^{46}$V & 1{*} & 0 & 0.0184 & 0.1523 & 6.73$\times$10$^{-4}$ & 0 & \tabularnewline
\hline 
\end{tabular}
\par\end{center}

\begin{center}
\textbf{Table III: log(ft) values}
\par\end{center}

\begin{center}
\begin{tabular}{|c|c|c|c|c|c|c|c|}
\hline 
Transition & $I_{i}$ & $I_{f}$ & MBZE & J=0 & J=Half & J=7 & Experiment\tabularnewline
\hline 
\hline 
$^{43}$Sc $\rightarrow^{43}$Ca & 3.5 & 2.5 & 4.519 & $\infty$ & 4.819 & 4.205 & 5.0\tabularnewline
\hline 
$^{43}$Sc $\rightarrow^{43}$Ca & 3.5 & 3.5 & 4.365 & 3.834 & 4.152 & 4.992 & 4.9\tabularnewline
\hline 
$^{43}$Sc $\rightarrow^{43}$Ca & 3.5 & 4.5 & 8.672 & $\infty$ & 7.074 & 6.458 & \tabularnewline
\hline 
$^{44}$Sc $\rightarrow^{44}$Ca & 2 & 2 & 4.888 & 4.804 & 5.440 & 5.178 & 5.3\tabularnewline
\hline 
$^{45}$Sc $\rightarrow^{45}$Ca & 3.5 & 2.5 & 5.619 & $\infty$ & 5.619 & 8.226 & \tabularnewline
\hline 
$^{45}$Ca $\rightarrow^{45}$Sc & 3.5 & 3.5 & 4.849 & 3.931 & 4.438 & 7.948 & \tabularnewline
\hline 
$^{45}$Sc $\rightarrow^{45}$Ca & 3.5 & 4.5 & 7.376 & $\infty$ & 7.092 & 8.578 & \tabularnewline
\hline 
$^{45}$Ti $\rightarrow^{45}$Sc & 3.5 & 3.5 & 4.425 & 4.415 & 4.353 & 6.821 & 4.6\tabularnewline
\hline 
$^{46}$Ti $\rightarrow^{46}$V & 4 & 4 & 5.779 & 5.370 & 4.130 & 7.284 & 6.2\tabularnewline
\hline 
$^{46}$Ti $\rightarrow^{46}$V & 4{*} & 4 & 5.828 & 3.854 & 6.336 & 6.788 & 6.2\tabularnewline
\hline 
$^{46}$Ti $\rightarrow^{46}$V & 1 & 0 & 4.694 & $\infty$ & 5.027 & 4.224 & 5.3\tabularnewline
\hline 
$^{46}$Ti $\rightarrow^{46}$V & 1{*} & 0 & 5.326 & 4.409 & 6.763 & $\infty$ & \tabularnewline
\hline 
\end{tabular}
\par\end{center}

Consider first the behaviour in going from J=0 pairing to J=7 pairing
via half and half. For the case $^{43}$Sc (I=7/2 T=1/2) $\rightarrow$
$^{43}$Ca (T=3/2) we find that when I$_{f}$ is 5/2 or 9/2, B(GT)
vanishes for J=0 pairing. For this interaction, seniority v is a good
quantum number. We can classify the states by (v,T,t) where t is the
reduced isospin. The initial I=7/2 state has v=1 and the final states
have v=3. The reduced isospins are also different, t=1/2 and t=3/2
respectively. It is not correct to say that seniority must be conseved
-- that is not the case. As discussed by Harper and Zamick {[}5,6{]},
with a J=0 pairing interaction one cannot have both the senority and
reduced isospin change at the same time.

As we go from J=0 pairing to J=7 pairing we get a steady increase
in B(GT) in the 7/2 $\rightarrow$ 9/2 and 7/2 $\rightarrow$ 5/2
cases. The former values are (0, 3.29$\times$10$^{-4}$, 0.00136)
whilst for 7/2 $\rightarrow$ 5/2 the values are (0, 0.0592, 0.2434).
We next consider 7/2 $\rightarrow$ 7/2 in $^{43}$Sc. Now we have
an opposite behaviour. The J=0 case yields the largest value for B(GT).

In $^{45}$Sc we have two examples of non-monotonic behaviour. This
is for the cases 7/2 $\rightarrow$ 9/2 and 7/2 $\rightarrow$ 5/2.
The 3 values are (0, 3.16$\times$10$^{-4}$, 1.03$\times$10$^{-5}$)
and (0, 9.4$\times$10$^{-3}$, 2.32$\times$10$^{-5}$) respectively.

In general, the values of B(GT) in $^{45}$Sc are smaller than in
$^{43}$Sc. It should be mentioned that systematics of B(GT)'s in
the f$_{7/2}$ region can be explained by the Lawson K selection rule
{[}7{]}.

We next carefully discuss the case I=1$^{+}$$\rightarrow$ I=0$^{+}$
in $^{46}$Ti. This was discussed by Harper and Zamick {[}6{]} but
in the context of an M1 transition B(M1). However, that makes no difference
because it was shown that B(GT) and the corresponding B(M1) were proportional.
There is, nonetheless, an apparent difference in the behaviour as
we go from J$_{max}$ pairing to J=0 pairing. Harper et al.{[}6{]}
state that there is non-monotonic behavour -- J=7 is relatively large,
half and half small, and J=0 pairing large again. But in the second
last row of the present work we get a monotonic decrease as we go
from J=7 to J=0.

The difference is that Harper et al. {[}6{]} always chose the state
of lowest energy whilst in the present work we take the state of lowest
energy for a fixed isospin. As we go to the J=0 pairing limit the
T=2 J=1$^{+}$ state in $^{46}$Ti state starts coming below a T=1
J=1$^{+}$ state. The B(GT) (or B(M1)) to the T=2 state is relatively
large and this explains why the value of B(GT), which first decreases
in going from J=7 to half and half, suddenly increases. If, as we
do in this work, we constrain the isospin to be unchanged, we get
the simpler monotonic behaviour. To get the Harper et al result {[}6{]}
we take the J=7 pairing and the half value from the second last row,
0.0307, and the J=0 result from the last row, 0.1532. The state 1$^{+}$
in this last row has isospin T=2, whereas in the second last row the
1$^{+}$ state is the lowest with T=1.

For B(GT) $^{46}$Ti 4 to 4 we have to take care since for J=0 pairing
the lowest 4$^{+}$ T=1 states are degenerate. We therefore slightly
remove the degeneracy by considering an interaction 0.9 J=0 pairing
and 0.1 J=7 pairing. We see that one of the B(GT)'s is small and the
other large. With MBZE the B(GT)'s to the lowest two 4$^{+}$ states
are both small. 

We next compare the 'realistic' MBZE results with experiment. Although
things are in the right ballpark, there are significant deviations,
indicating the need for configuration mixing.

\section{Magnetic moments}

In table IV we show a corresponding study of magnetic moments.

\begin{center}
\textbf{Table IV: Magnetic moments}
\par\end{center}

\begin{center}
\begin{tabular}{|c|c|c|c|c|c|c|}
\hline 
Nucleus & Spin & MBZE & J=0 & J=Half & J=7 & Experiment\tabularnewline
\hline 
\hline 
$^{43}$Sc & 3.5 & 4.324 & 3.614 & 4.204 & 4.328 & +4.62\tabularnewline
\hline 
$^{44}$Sc & 2 & 1.990 & 0.592 & 1.779 & 2.268 & +2.56\tabularnewline
\hline 
$^{45}$Sc & 3.5 & 4.646 & 4.468 & 4.703 & 4.158 & +4.76\tabularnewline
\hline 
$^{45}$Ti & 2.5 & -0.764 & 0.041 & -0.905 & -0.751 & -0.133\tabularnewline
\hline 
$^{45}$Ti & 3.5 & -0.604 & -0.891 & -0.779 & -0.377 & 0.095\tabularnewline
\hline 
$^{46}$Ti & 2 & 0.991 & 1.990 & 1.152 & 0.613 & -0.98\tabularnewline
\hline 
\end{tabular}
\par\end{center}

It should be noted that since 1964 a new magnetic moment has been
measured experimentally -- that of $^{45}$Ti. The value is 0.095,
but the sign is undetermined. All our interactions yield negative
magnetic moments. The closest is the case of J$_{max}$ pairing which
gives -0.377, still a big discrepancy. 

It should be noted that Ricardo Garcia has two institutional affiliations:
Rutgers University, and the University of Puerto Rico, Rio Piedras
Campus. The permament address associated with the UPR-RP is University
of Puerto Rico, San Juan, Puerto Rico 00931. RG acknowledges that
to carry out this work he has received support from the U.S. National
Science Foundation through grant PHY-1263280, and thanks the REU Physics
program at Rutgers University for their support.

\end{document}